\documentclass{aa}
\usepackage{graphicx}
\usepackage{txfonts}
\usepackage{natbib}
\bibpunct{(}{)}{;}{a}{}{,}
\begin{document}

\title{Physical properties of two compact high-velocity clouds possibly associated with the Leading Arm of the Magellanic System\thanks{The Australia Telescope Compact Array and the Parkes telescope are part of the Australia Telescope which is funded by the Commonwealth of Australia for operation as a National Facility managed by CSIRO.}}
\titlerunning{Properties of two compact high-velocity clouds}
\subtitle{}
\author{N. Ben Bekhti
    \and
        C. Br\"{u}ns
    \and
        J. Kerp
    \and
        T. Westmeier}
\offprints{N. Ben Bekhti}
\institute{Argelander-Institut f\"ur Astronomie\thanks{Founded by merging of the Institut f\"ur Astrophysik und Extraterrestrische Forschung, the Sternwarte, and the Radioastronomisches Institut der Universit\"at Bonn.}, Universit\"{a}t Bonn, Auf dem H\"{u}gel 71, 53121 Bonn, Germany\\
\email{nbekhti@astro.uni-bonn.de}
}
\date{Received month ??, ????; accepted month ??, ????}

\abstract{}{We observed two compact high-velocity clouds HVC~291+26+195 and HVC~297+09+253 to analyse their structure, dynamics, and physical parameters. In both cases there is evidence for an association with the Leading Arm of the Magellanic Clouds. The goal of our study is to learn more about the origin of the two CHVCs and to use them as probes for the structure and evolution of the Leading Arm.}
{We have used the Parkes 64-m radio telescope and the Australia Telescope Compact Array (ATCA) to study the two CHVCs in the 21-cm line emission of neutral hydrogen. The observations with a single-dish and a synthesis telescope allow us to analyse both the diffuse, extended emission as well as the small-scale structure of the clouds. We present a method to estimate the distance of the two CHVCs.}{The investigation of the line profiles of HVC~297+09+253 reveals the presence of two line components in the spectra which can be identified with a cold and a warm gas phase. In addition, we find a distinct head-tail structure in combination with a radial velocity gradient along the tail, suggesting a ram-pressure interaction of this cloud with an ambient medium. HVC~291+26+195 has only a cold gas phase and no head-tail structure. The ATCA data show several cold, compact clumps in both clouds which, in the case of HVC~297+09+253, are embedded in the warm, diffuse envelope. All these clumps have very narrow \ion{H}{i} lines with typical line widths between $2$ and $4 \; \mathrm{km \, s}^{-1}$ FWHM, yielding an upper limit for the kinetic temperature of the gas of $T_{\mathrm {max}} = 300 \; \mathrm{K}$. We obtain distance estimates for both CHVCs of the order of 10 to $60 \; \mathrm{kpc}$, providing additional evidence for an association of the clouds with the Leading Arm. Assuming a distance of $50 \; \mathrm{kpc}$, we get \ion{H}{i} masses of $5.9 \cdot 10^3 \; \textit{M}_{\odot}$ and $4.0 \cdot 10^4 \; \textit{M}_{\odot}$ for HVC~291+26+195 and HVC~297+09+253, respectively.}{}


\maketitle

\section{Introduction}

High-velocity clouds \citep[HVC,][] {mulleroortraimond63} are clouds of neutral atomic hydrogen, with radial velocities ($|v_\mathrm{LSR}| \gtrsim 100$ \textrm{km\,s}$^{-1}$) inconsistent with a simple model of galactic rotation. They are believed to be extraplanar objects located in the halo \citep[e.g. complex A, ][]{vanWoerden99} and circumgalactic environment of disk galaxies \citep[e.g. the Magellanic Stream, ][]{mathewsonclearymurray74}. \citet{Wakker91} introduced the so-called deviation velocity, $v_\mathrm{dev}$, that is
the difference between the velocity of the clouds in the LSR frame and the extreme velocity
value consistent with the Milky Way gas distributed along the line of sight.
A cloud is defined as an HVC, if $|v_\mathrm{dev}| \geq 50$ \textrm{km\,s}$^{-1}$.

HVCs can be detected all over the sky, but they are not homogeneously distributed \citep{Wakkervanwoerden97,Murphy95,deheijbraunburton02}.
On the one hand, there are coherent and extended objects like the complexes A, C and M and the Magellanic Stream. On the
other hand, there are compact, isolated clouds with angular diameters of $\varphi \leq 2\degr$ FWHM, the
so-called compact high-velocity clouds (CHVCs), which are kinematically and spatially separated from the gas
distribution in their environment \citep{braunburton99}.

The most critical issue of HVC research is the determination of distances. It is very difficult to find suitable
background sources at known distances against which the clouds would appear in absorption. Due to this fact the
spatial distribution of HVCs is largely unknown, and important distance-dependent physical parameters, like mass
($M \sim d^2$), radius ($R \sim d$) and particle density ($n \sim d^{-1}$), are poorly constrained. The
distances of only three complexes have been determined so far via absorption line measurements. For complex M a
distance of $1.5 < z < 4.4$ kpc was obtained by \citet{Danly93}, and a distance bracket for complex A of
$4 < z < 10$ kpc was determined by \citet{vanWoerden99}. Recently, \citet{thom2006} derived a distance bracket for HVC complex WB of $7.7 < d < 8.8$ kpc.

There are three main hypotheses for the origin of HVCs. The first is the formation from gas flowing out of
the Galactic disk \citep[galactic fountain model, ][]{shapirofield76, bregman80}, the second is the origin from the interaction of dwarf galaxies
with the Milky Way \citep{gardinernoguchi96, blandhawthorn98, bregman04}, and the third hypotheses is the primordial gas model, where HVCs represent intergalactic
gas which was not yet accreted by one of the galaxies in the Local Group \citep{oort66, Blitzetal99}.

Several compact high-velocity clouds (CHVCs) have been studied in detail \citep{braunburton00, burtonbraunchengalur01, westmeier05}. In many cases, two-component
line profiles are observed which can be identified with a warm and a cold gas phase \citep{Giovanelli73,Greisen76,wolfiremckeehollenbachtielens95}. Furthermore, some clouds
show a head-tail structure or horse-shoe shape connected with gradients in column density and in
line width across the symmetry axis. These observations suggest an interaction of the clouds with
their ambient medium. Interferometric observations \citep{braunburton00,deheijbraunburton02}
revealed that these clouds have a characteristic morphology. Compact cores are embedded in diffuse warm \ion{H}{i} gas. The clumpy substructures have relatively high column densities ($\approx 10^{20}$ cm$^{-2}$)
and very small line widths down to 2 \textrm{km\,s}$^{-1}$ FWHM. This indicates that the compact cores can be identified with a cool neutral medium.

Recently, a statistical analysis of the high-velocity sky was carried out by \citet{kalberlahaud06} based on the Leiden/Argentine/Bonn (LAB) Galactic \ion{H}{i} Survey \citep{kalberlaetal05}. They searched all HVC complexes for indications of multi-phase structures, excluding CHVCs which are not sufficiently resolved by the LAB Survey. A multi-phase structure was found in $f_{\rm m} = 23.8$\% of all sight lines investigated by \citet{kalberlahaud06}, corresponding to a contribution of the cold cores to the total \ion{H}{i} column density of $f_{N} = 20.8$\%. 
Thus, a multi-phase structure is a common phenomenon in high-velocity clouds which means that many HVCs contain cold, compact cores embedded in a diffuse envelope of warm neutral gas. Two-component line profiles are also frequent in the Magellanic System. In the Leading Arm $f_{\rm m} = 23.8$\% of sight lines with HVC emssion show a multi-phase structure, equivalent to a column density contribution of the cold gas of $f_{N} = 13.5$\%. The corresponding numbers for the Magellanic Stream are $f_{\rm m} = 38.0$\% and $f_{N} = 27.0$\% for the region with positive radial velocities but only $f_{\rm m} = 7.5$\% and $f_{N} = 11.0$\% for the negative-velocity region. The multi-component structure of the Leading Arm is confirmed by \ion{H}{i} observations with the Parkes telescope and the ATCA by \citet{Bruenskerp05} and \citet{bruenswestmeier04}.

\citet{wakkeroosterlooputman02} studied a denser region in one of the filaments of the Leading Arm
of the Magellanic System using the Parkes telescope and the ATCA. Their data reveal a very complex internal structure
of the high-velocity gas. The filamentary structure introduces a high degree of complexity which does not allow to draw conclusions about the physical conditions in the Leading Arm and its environment. 

\begin{figure}
\includegraphics[width=0.48\textwidth,clip]{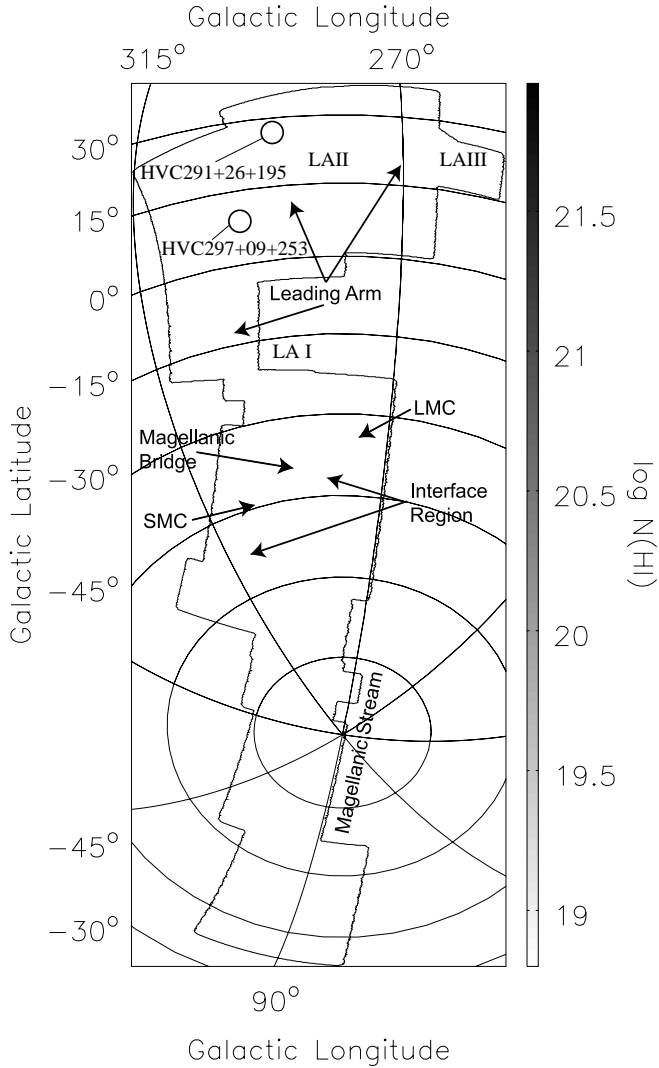}
\caption{\ion{H}{i} column density distribution of the Magellanic System from \citet{Bruenskerp05}. The column densities range from $N_\mathrm{HI}=6 \cdot 10^{18}$ cm$^{-2}$ (light grey) to $N_\mathrm{HI}=9 \cdot 10^{21}$ cm$^{-2}$ (black). The black circles mark the positions of HVC~297+09+253  and HVC~291+26+195.}
\label{fig_MS_NHImap}
\end{figure}

We decided to study two compact clouds, HVC~291$+$26$+$195 and HVC~297$+$09$+$253, which are isolated but located in the vicinity of the Leading Arm. Fig.~\ref{fig_MS_NHImap} shows a column density map of the Magellanic System \citep{Bruenskerp05}. The two circles mark the positions of the two CHVCs. Both clouds were observed with the Parkes telescope and the ATCA. The observations with both a single-dish telescope and an interferometer allow us to analyse the total
\ion{H}{i} mass and the extended structure as well as the small-scale structure within the clouds. These clouds are supposed to have a relatively simple structure, allowing us to uncover their physical state and that of the ambient medium. If the CHVCs were associated with the Leading Arm we would be able to roughly constrain their distance and origin, which are crucial but unknown for most HVCs/CHVCs. This would allow us to determine distance-dependent physical parameters such as mass and density and to study the interaction effects observed in connection with the Leading Arm. Furthermore, with the high-resolution synthesis observations we are able to investigate a possible multi-component structure in both clouds which were not included in the sample of \citet{kalberlahaud06} since the spatial resolution of the LAB Survey is not adequate to resolve them.

Our paper is organised as follows.
In Sect.~\ref{data} we describe the data acquisition and data reduction. In Sect.~\ref{results} the
results of the analysis of the Parkes and the ATCA data of both CHVCs are presented. In Sect.~\ref{disscussion}
we discuss the results regarding ram-pressure interaction, a distance estimate and a probable association
with the Leading Arm of the Magellanic System.
Sect. \ref{summary} summarises our results and point out the importance of follow-up observations that will
improve our understanding of the nature of high-velocity clouds.

\section{Data acquisition and reduction}\label{data}

\subsection{Parkes data}

The Parkes data were obtained as part of an \ion{H}{i} survey of the Magellanic System \citep{Bruenskerp05}.
The Parkes telescope has a diameter of 64 m. At 21 cm wavelength the half-power beam width (HPBW) is 14$\farcm$1.
For our observation the velocity resolution is about 1 \textrm{km\,s}$^{-1}$. The width of a spectral bin corresponds to 0.825
\textrm{km\,s}$^{-1}$  using the Parkes 2048-channel autocorrelator. The \ion{H}{i} data were observed in on-the-fly mode.
The integration time per spectrum was 5 s and the beam moved $5\arcmin$ on the sky during the integration
\citep{Bruenskerp05}.

The single-dish data were analysed with the GILDAS software CLASS. In all spectra polynomial baselines up to
$5^\mathrm{th}$ order were fitted. For this purpose windows were set individually around the line emission. The data within
these windows were not considered for the fit. After the baseline fit Gaussian functions were fitted in the spectral lines. The criterion
for line emission was set to 3$\sigma_\mathrm{rms}$ at a velocity resolution of 2 \textrm{km\,s}$^{-1}$.
The noise of the Parkes data is about $\sigma_\mathrm{rms}=0.1$~K. For the further data analysis data cubes for
both clouds were prepared.

\subsection{ATCA data}

The ATCA is an east-west interferometer with six
antennas. Each antenna has a diameter of 22~m. Of these six antennas only five were used for the following
analysis. The ATCA data were observed using the 750D configuration. For this configuration the five antennas provide baselines between 30 m and 720 m. The baselines associated with the sixth
antenna correspond to very small angular scales, where no signal is detected. For our observations we chose a correlator
with a bandwidth of 4~MHz and a velocity resolution of about 0.8~\textrm{km\,s}$^{-1}$. The width of a
spectral channel is 0.8~\textrm{km\,s}$^{-1}$  with a total of 1024~channels. The observing time for each cloud
was 12.5~hours.

The ATCA data were analysed with the MIRIAD software. The source 1934--638 was used as primary calibrator for
the flux calibration. The primary calibrator was observed for about 15 minutes each at the beginning and at
the end of the observation. The secondary calibrator, 1215--457, was used for the gain and bandpass calibration,
this source was observed once every hour for 5 minutes. The pointing centre is $\alpha(\textrm{J}2000)=12^\mathrm{h} 12^\mathrm{m} 35^\mathrm{s}$, $\delta(\textrm{J}2000)=-53\degr 42\arcmin 30\arcsec$ for HVC\,297+09+253 and $\alpha(\textrm{J}2000)=11^\mathrm{h} 58^\mathrm{m} 14^\mathrm{s}$, $\delta(\textrm{J}2000)=-35\degr 34\arcmin 36\arcsec$ for HVC\,291+26+195. We used a robustness parameter of 0.5 to produce a data
cube. This robustness parameter is a compromise between a good resolution and a high signal-to-noise ratio (SNR).
The image size of one plane of the data cube is $300 \times 300$ pixels, the pixel size is $8\arcsec$. The HPBW
of the elliptical beam for the corresponding parameters is about $58\arcsec \times 36\arcsec$ for HVC\,297+09+253
and $75\arcsec \times 38\arcsec$ for HVC\,291+26+195. The deconvolution was performed with the CLEAN algorithm \citep{hoegbom74}.
The final data cube has an rms of about $7.6$~mJy~beam$^{-1}$ towards the centre of the field
which corresponds to 1.7~K.

\section{Results}\label{results}

\subsection{HVC\,297+09+253}\label{hvc1}

\subsubsection{Parkes data}

\begin{figure*}
\includegraphics[width=\textwidth]{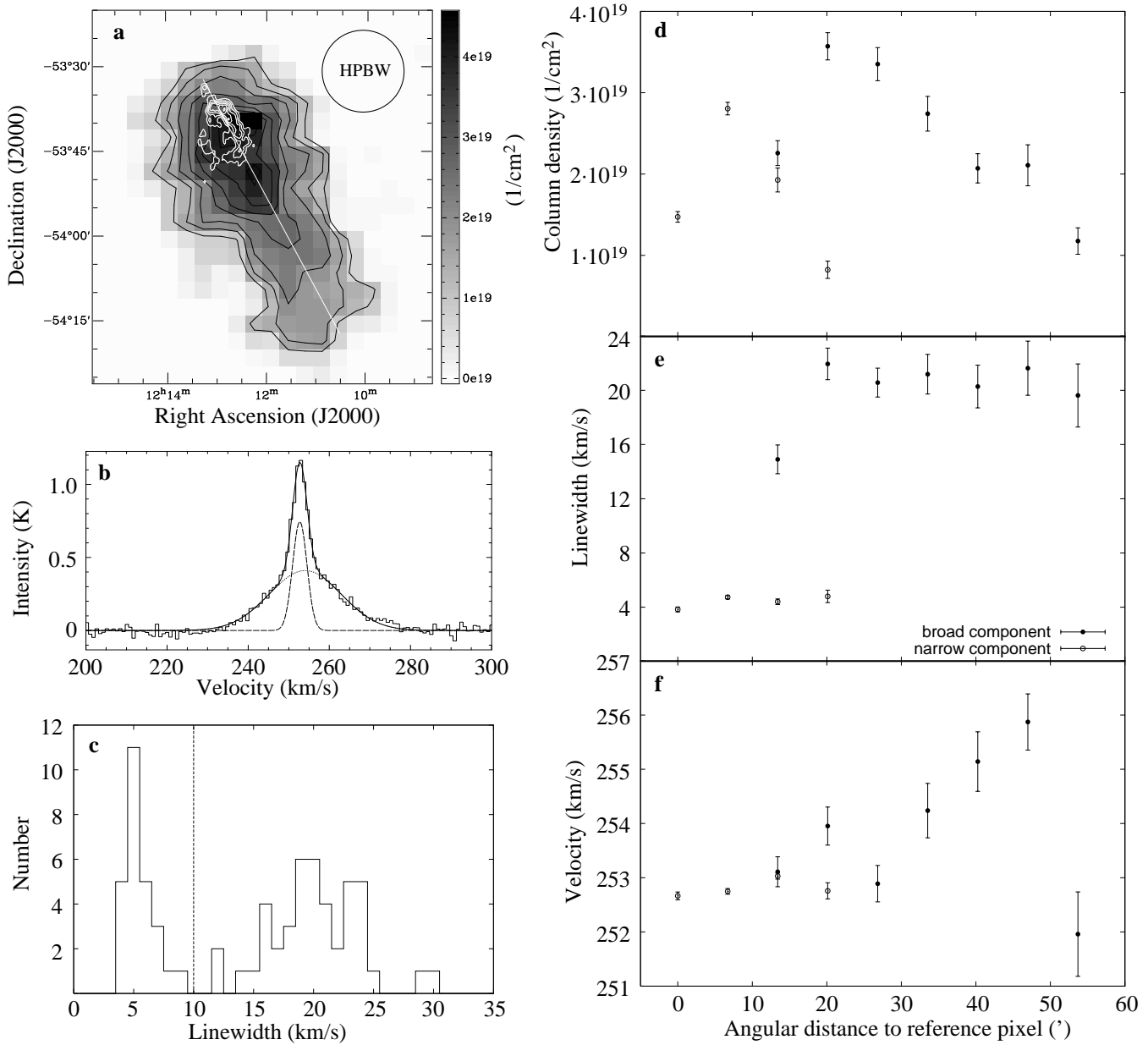}
\caption{
\textbf{a}: \ion{H}{i} column density map of HVC\,297+09+253 observed with the Parkes
telescope. The angular resolution of HPBW = 14$\farcm$1 is indicated in the upper right.
The black contours range from $1 \cdot 10^{19}$ to $4 \cdot 10^{19}$ cm$^{-2}$ in steps of
$5 \cdot 10^{18}$ cm$^{-2}$. Overlaid are the contour lines (white) of this cloud observed with the
ATCA ($5 \cdot 10^{19}$ to $3 \cdot 10^{20}$ cm$^{-2}$ in steps of $5 \cdot 10^{19}$ cm$^{-2}$).
The ATCA data cover the northern half of the cloud. The white line indicates the major axis of
HVC\,297+09+253.
\textbf{b}: The average of all 59 spectra with $T_\mathrm{B}> 3 \sigma_\mathrm{rms}$
observed with Parkes. Apparently, there is a superposition of a narrow and a broad line component.
We fitted two Gaussian functions into the spectrum. The sum of the two fits is also shown (solid curve).
The resulting fit reproduces the line profile very well.
\textbf{c}: The histogram shows the distribution of line widths of HVC\,297+09+253. Two separate line
components are present. A line width of \mbox{$\Delta v_\mathrm{FWHM}=10$ \textrm{km\,s}$^{-1}$ FWHM}
(dashed line) can be used to separate the two line components.
\textbf{d-f}: Column density, $N_\mathrm{HI}$, linewidth, $\Delta v_\mathrm{FWHM}$, and
mean velocity, $v_\mathrm{LSR}$, from the two-component Gaussian fits along the major axis of HVC\,297+09+253
(see Fig.~\ref{hvc297composite}a). The diagrams demonstrate that the cool and the warm gas phase are partly spatially separated.}
\label{hvc297composite}
\end{figure*}

Fig. \ref{hvc297composite}a shows a column density map of the Parkes data of HVC\,297+09+253.
The CHVC shows an elongated head-tail structure with a major axis of about 1$^\circ$ and a minor axis that
is not resolved by the Parkes beam.

The average of all 59 spectra of the CHVC reveals a two-component line structure, where a narrow line
component is superposed on a broad line component (Fig. \ref{hvc297composite}b). The figure also shows the results
of a two-component Gaussian fit to the average spectrum.
The line profile is well represented by two Gaussians. The Gaussian fits provide line widths of
$20.9 \pm 0.6$ \textrm{km\,s}$^{-1}$  FWHM for the broad component and $4.2 \pm 0.2$ \textrm{km\,s}$^{-1}$  FWHM
for the narrow component. The so-called Doppler temperature assumes a pure Maxwellian velocity distribution and provides
an upper temperature limit for the gas
\begin{equation}
\frac{T_\mathrm{D}}{[\mathrm{K}]}=21.8 \left(\frac{\Delta v_\mathrm{FWHM}}{[\textrm{km\,s}^{-1}]}\right)^2,
\end{equation}
where $\Delta v_\mathrm{FWHM}$ is the line width. The resulting upper temperature limits are
$9500 \pm 550$ K (broad component) and $380 \pm 40$ K (narrow component). The kinetic temperatures of the two gas
phases are expected to be lower, as e.g. turbulent motions also contribute to the observed line width.

For a more detailed analysis of the physical parameters of the two gas phases Gaussian fits were performed for
the individual spectra of HVC\,297+09+253. Two Gaussian components were fitted to those spectra that are not well
represented by a single Gaussian. Fig. \ref{hvc297composite}c shows a histogram of the line widths resulting from
the Gaussian decomposition.
There are two separate distributions with maxima at 5 \textrm{km\,s}$^{-1}$  FWHM for the narrow component and
about 20 \textrm{km\,s}$^{-1}$  FWHM for the broad line component. The histogram implies that a line width of about 10 \textrm{km\,s}$^{-1}$ provides an adequate separation between the two gas components. This line width corresponds to a temperature of 2000 K.

\begin{figure}
\includegraphics[width=0.46\textwidth,clip]{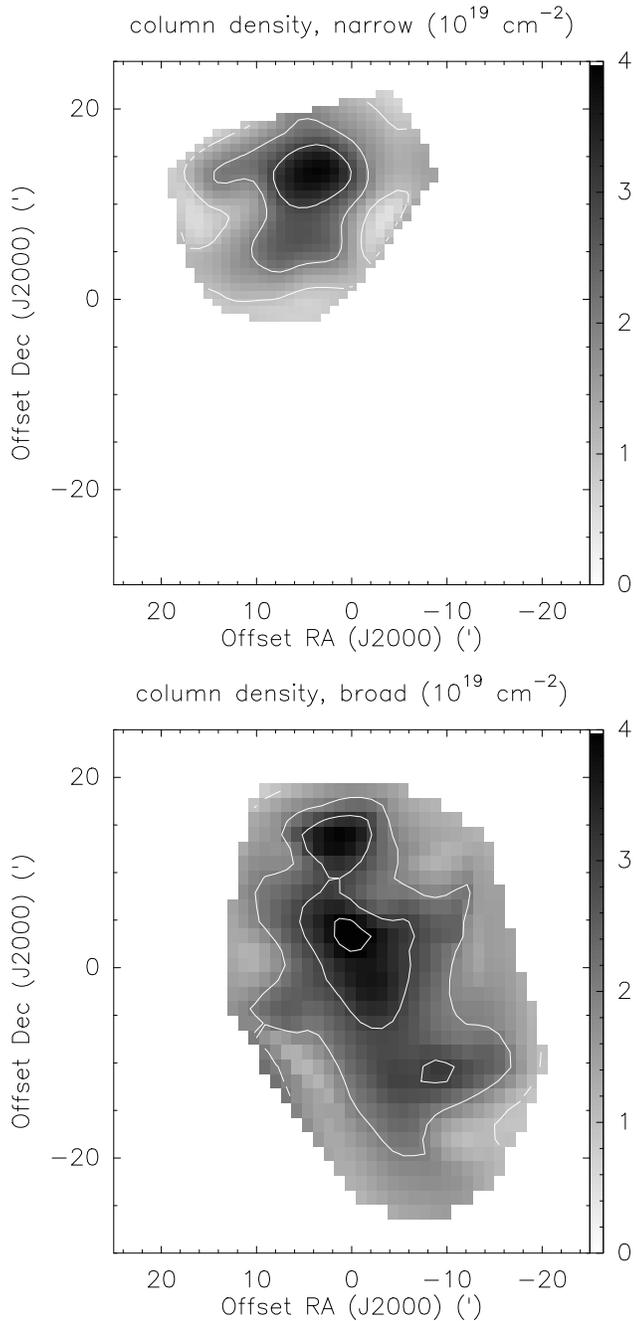}
\caption{The spatial distribution of the column densities of both line components of HVC~297+09+253.
\textbf{Upper panel}: Column density of the narrow line component. \textbf{Lower panel}: Column density of the broad component.
The contours range from $1 \cdot 10^{19}$ to $4 \cdot 10^{19}$~cm$^{-2}$ in steps of $1 \cdot 10^{19}$~cm$^{-2}$.}
\label{fig_hvc1_linewidth}
\end{figure}

The distribution of the parameters $N_\mathrm{HI}$, $v_\mathrm{LSR}$ and $\Delta v_\mathrm{FWHM}$ is analysed along
the major axis of HVC\,297+09+253  (white line in Fig. \ref{hvc297composite}a), and the results are shown in
Fig.~\ref{hvc297composite}d-f. The diagrams demonstrate that the two gas phases are partly spatially separated. The maxima of the
column density of the two line components are displaced by $\approx 15\arcmin$. The narrow component is very compact with
a FWHM diameter of only $\approx 15 \arcmin$ which is similar to the Parkes HPBW of $14\farcm1$, indicating that the cold component
is not resolved by the Parkes telescope. The column density of the warm component rises steeply close to the cold component
and decreases slowly towards the southern end of HVC\,297+09+253.
The spatial distribution of the column density of the two gas phases is also shown in Fig.~\ref{fig_hvc1_linewidth}.
The peak column densities are $N_\mathrm{HI}=(3.0 \pm 0.1)  \cdot 10^{19} $ cm$^{-2}$ and
$N_\mathrm{HI}=(3.9 \pm 0.2)  \cdot 10^{19} $ cm$^{-2}$ for the cold and warm gas phase, respectively.
The cold gas phase is only detected in the northern part of the cloud, whereas the warm gas phase extends over the
entire cloud.

Fig.~\ref{hvc297composite}e shows the line widths along the major axis of HVC\,297+09+253. The line widths of the cold and warm
component are approximately 5 \textrm{km\,s}$^{-1}$ and 20 \textrm{km\,s}$^{-1}$ FWHM, respectively, without much
variation over the extent of the cloud.

The radial velocity (LSR frame) is plotted in Fig. \ref{hvc297composite}f. The cold gas phase has $v_{\rm LSR}\approx$
253 \textrm{km\,s}$^{-1}$. The warm component has $v_{\rm LSR}\approx$ 253 \textrm{km\,s}$^{-1}$ close to the position of
the cold component and increases up to 256 \textrm{km\,s}$^{-1}$  towards the southern end of HVC\,297+09+253.
This velocity gradient will be discussed in Sect. \ref{LeadingArm}.

\begin{figure}
\includegraphics[width=0.45\textwidth,clip]{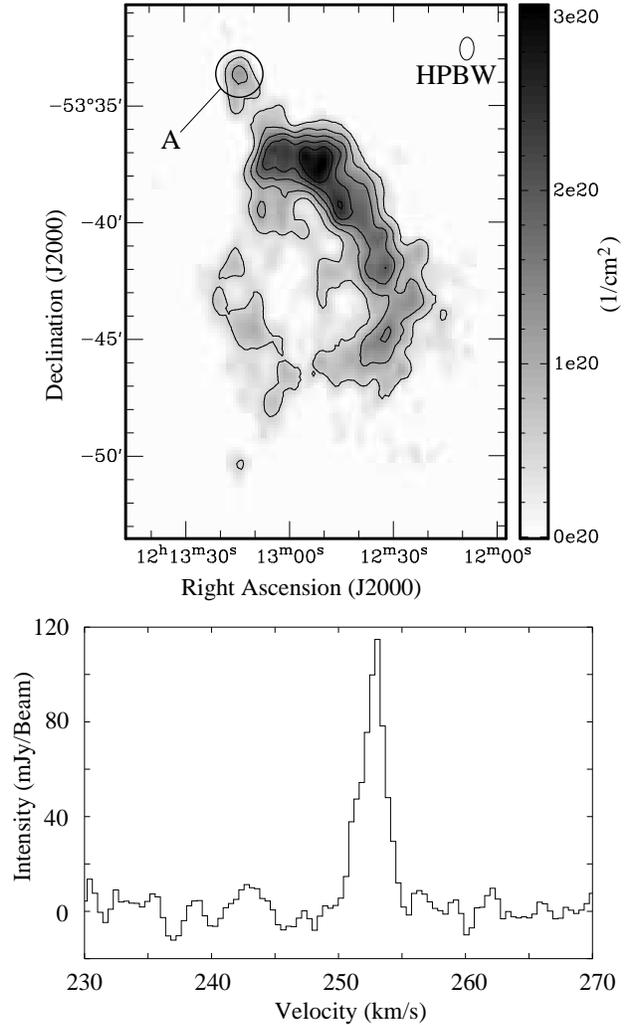}
\caption{\textbf{Upper panel}: Column density map of the ATCA data cube of HVC\,297+09+253. The black circle marks the position of clump A, which is analysed in detail regarding its physical properties. The contours range from $5 \cdot 10^{19}$ to $3 \cdot 10^{20}$ cm$^{-2}$ in steps of $5 \cdot 10^{19}$ cm$^{-2}$. The ATCA beam of $58.6\arcsec \times 36.2\arcsec$ HPBW is also shown. \textbf{Lower panel}: A typical spectrum in the direction of clump A of HVC\,297+09+253.}
\label{fig_hvc1_ATCA_clump_specA}
\end{figure}

\subsubsection{ATCA data}\label{hvc1atca}

The ATCA observations of HVC\,297+09+253  were obtained towards the high column density region, the so-called head, of the cloud,
where the cold gas phase is found. Fig.~\ref{fig_hvc1_ATCA_clump_specA} shows the column density of HVC\,297+09+253 as
observed with the ATCA. For comparison, Fig.~\ref{hvc297composite}a shows the column density map of HVC\,297+09+253 observed
with Parkes superposed by the ATCA observations as white contours.

The total \ion{H}{i} mass is determined by
\begin{equation}
M_\mathrm{HI}=m_\mathrm{H} d^2 \tan^2 \varphi \sum_i N_\mathrm{HI}^{(i)}
\end{equation}
where $m_\mathrm{H}$ is the mass of a hydrogen atom, $d$ is the distance to the cloud, $\varphi$ is the angular size of a single
pixel of the column density map, and $N_\mathrm{HI}$ is the column density. The total \ion{H}{i} mass detected by the ATCA
is $M_\ion{H}{i}$ = $1.7 \cdot 10^4~\textit{M}_{\sun} \,[\textit{d}/50\,\mathrm{kpc}]^2$, while the total mass detected with Parkes is
$M_\ion{H}{i}$ = $4.7 \cdot 10^4~\textit{M}_{\sun}\,[\textit{d}/50\,\mathrm{kpc}]^2$.
We determine a lower detection limit $M_\mathrm{limit}$ by computing the mass of a point source with a typical line width which generates a $3\sigma_\mathrm{rms}$ (Parkes) and $5 \sigma_\mathrm{rms}$ (ATCA) signal in the mass map. For a typical line width of $\Delta v_\mathrm{FWHM}= 4$ $\mathrm{km \, s}^{-1}$ we get $M_\mathrm{limit}=600~\textit{M}_{\sun}$ (Parkes) and $M_\mathrm{limit}=40~\textit{M}_{\sun}$ (ATCA).
A comparison of the total \ion{H}{i} masses obtained for the Parkes and ATCA data for HVC\,297+09+253 reveals that the
ATCA detects only $36 \%$ of the mass of HVC\,297+09+253. This result demonstrates that the diffuse extended gaseous phase,
constituting a major part of the total \ion{H}{i} mass, is missed by the interferometer.

A combination of the single-dish and the interferometer data is nevertheless not meaningful. While the warm gas phase
has considerable column densities, the line intensities are quite low due to the large line widths.
The rms noise of the ATCA data ($\sigma_\mathrm{rms} \approx 1.7$ K) is significantly larger than typical line intensities of
the warm component, which are well below 1~K (the noise of the Parkes data of $\sigma_\mathrm{rms} \approx 0.1$ K is much lower).

The high resolution ATCA data cube reveals a clumpy substructure of HVC\,297+09+253 (see Fig.~\ref{fig_hvc1_ATCA_clump_specA}).
Gaussian functions were fitted to the spectra of all clumps to analyse their physical properties in detail.
All clumps in HVC\,297+09+253 have very small line widths between 2 and 4 \textrm{km\,s}$^{-1}$  FWHM. The corresponding upper limit for the kinetic
temperature is $T_\mathrm{kin} \leq 300$ K.
The clumps have relatively high column densities of $N_\mathrm{HI} \approx 10^{20}$ cm$^{-2}$ and small angular sizes of
$\varphi \approx 1\arcmin \ldots 2\arcmin$ FWHM. The physical parameters are shown in Table \ref{tab_hvc_atca}.

The position of one exemplary core, clump~A, is marked in Fig.~\ref{fig_hvc1_ATCA_clump_specA}.
Clump~A has a column density of $N_\mathrm{HI}=(1.3 \pm 0.1) \cdot 10^{20}$ cm$^{-2}$.
The spectrum of clump A (see Fig. \ref{fig_hvc1_ATCA_clump_specA}) has a very small line width of
$\Delta v_\mathrm{FWHM}=2.3 \pm 0.1$ \textrm{km\,s}$^{-1}$, yielding an upper temperature limit of $T_\mathrm{D}=119 \pm 13$ K.

The radiative transfer equation for the 21 cm line emission in the case without background emission is
\begin{equation}\label{equ_tau}
T_\mathrm{B}=T_\mathrm{spin}(1 - \mathrm{e}^{-\tau}),
\end{equation}
where $T_\mathrm{spin}$ is the spin temperature and $\tau$ the optical depth. The equation shows that
$T_\mathrm{B} \leq T_\mathrm{spin}$. In the case of a thermal excitation it is expected that
$T_\mathrm{spin} \approx T_\mathrm{kin}$. The measured brightness temperature $T_\mathrm{B}$ is therefore
a lower temperature limit, leading to a possible temperature range of
$T_\mathrm{B} \leq T_\mathrm{kin} \leq T_\mathrm{D}$ of the \ion{H}{i} gas.
The brightness temperature for clump A is $T_\mathrm{B}=32 \pm 4$ K. The very small line width allows us to
constrain the kinetic temperature of this clump to $32~\mathrm{K}\leq T_\mathrm{kin} \leq 120$ K.

Furthermore, Eq.~\ref{equ_tau} allows us to estimate a lower limit for the optical depth of clump A of
$\tau \ge 0.31 \pm 0.05$. This is a remarkable result because the approximation $\tau \ll 1$ (which is the common assumption to derive the column density as a linear function of $T_\mathrm{B}$) is not satisfied anymore. As we have only lower limits for the optical depth, we did not correct the column densities, which accordingly represent lower
limits.

The neutral atomic hydrogen density of the condensation can be estimated assuming a spherical symmetry and a constant density
$n=\frac{N_\mathrm{HI}}{\tan \phi d}$, where $N_\mathrm{HI}$ is the column density, $\phi$ is the angular diameter of the clump,
and $d$ is the distance to the cloud.
Condensation A is sufficiently isolated and circular symmetric to expect that its extent along the line of sight is similar to its apparent diameter in the plane of the sky. The estimated \ion{H}{i} volume density for condensation A is $n= 2.1\, \mathrm{cm}^{-3}~[d/\,50~\mathrm{kpc}]^{-1}$. Table \ref{tab_hvc_atca} compiles the estimated densities of
all condensations of HVC\,297+09+253. Note, however, that most of them are embedded in an enveloping medium making the estimates
less reliable. Furthermore we cannot rule out that the cold clumps contain substructure at sub-arcmin scale which is not resolved with our ATCA data. Consequently, the filling factor may be less than unity. Potential gaps between the dense clumps may be filled with warm neutral gas, but we can rule out the presence of molecular or highly ionised gas. Even the densest clumps have densities which are far too low to allow for the formation of molecules. There is not even a reliable indication of the existence of dust in HVCs \citep[e.g.][]{wakkerboulanger86}. Highly ionised gas also cannot exist in the inner parts of HVCs because they are shielded against the extragalactic radiation field by the surrounding warm neutral medium. For example, \citet{sternbergmckeewolfire02} perform hydrostatic simulations of CHVCs embedded in an ionised intergalactic medium. In their model the ionisation rate of hydrogen drops by several orders of magnitude from $1 \cdot 10^{-14}~\mathrm{s}^{-1}$ to $1 \cdot 10^{-17}~\mathrm{s}^{-1}$ from the outer edge to the centre of the cloud. In addition, the diffusion of hot plasma from the surrounding medium into the neutral cloud will be suppressed by a magnetic field in the intergalactic medium due to a magnetic barrier of increased field strength building up at the interface of the HVC \citep{Konz}

The pressure of the condensations can be estimated using the ideal gas equation,
\begin{equation}
\frac{p}{\rm{k_\mathrm{B}}}=\frac{N_\mathrm{HI}}{d \tan \phi} T,
\end{equation}
where $p$ is the pressure and $\rm{k_\mathrm{B}}$ the Boltzmann constant. We derive a relatively narrow range of allowed
pressures for condensation A of $(p / \rm{k_\mathrm{B}})_\mathrm{min} = 67 \pm 15$~K~cm$^{-3}~[d/\,50~\mathrm{kpc}]^{-1}$ and
$(p / \rm{k_\mathrm{B}})_\mathrm{max} = 256 \pm 53$~K~cm$^{-3}~[d/\,50~ \mathrm{kpc}]^{-1}$. The distance of HVC\,297+09+253 is
discussed in Sect. \ref{distance} and \ref{LeadingArm}. Table \ref{tab_hvc_atca} shows the distance-dependent parameters size,
mass, particle number, density, and pressure for all clumps of HVC\,297+09+253.

\begin{table*}
\caption{Important physical parameters for all clumps of HVC 297+09+253. For all distance-dependent parameters a distance of 50~kpc is assumed. The columns give the name of the clump, the coordinates, the velocity in the LSR frame, the line width $\Delta v_\mathrm{FWHM}$, the lower and upper temperature limits $T_\mathrm{min}$ and $T_\mathrm{max}$, the SNR, the optical depth $\tau$, the column density $N_\mathrm{HI}$, the angular diameter $\phi$, the particle density $n$, and the lower $\left(\frac{p}{k_\mathrm{B}}\right)_\mathrm{min}$ and upper pressure limit $\left(\frac{p}{k_\mathrm{B}}\right)_\mathrm{max}$ as well as the estimated distances $d$.}
\label{tab_hvc_atca}
\tiny
\centering
\begin{tabular}{ccccccccccccccc}\hline\hline
\rule{0pt}{3ex}clump&RA&Dec&$v_\mathrm{LSR}$&$\Delta v_\mathrm{FWHM}$&$T_\mathrm{max}$&$T_\mathrm{min}$&SNR&$\tau_\mathrm{min}$&$N_\mathrm{HI}$&$\phi$&$n$&$\left(\frac{p}{k}\right)_\mathrm{min}$&$\left(\frac{p}{k}\right)_\mathrm{max}$&$d$\\

\rule{0pt}{3ex}&&&[km\,s$^{-1}$]&[km\,s$^{-1}$]&[K]&[K]&&&[cm$^{-2}$]&[$\arcsec$]&[cm$^{-3}$]&[cm$^{-3}$K]&[cm$^{-3}$K]&[kpc]\\[1ex]\hline
\rule{0pt}{3ex}A & 12$^\mathrm{h}$13$^\mathrm{m}$14.5$^\mathrm{s}$& $-$53$\degr$33$\arcmin$32.8$\arcsec$ & 252.8  & 2.3  & 119 & 32  & 16.4 & 0.27 & $1.3\cdot10^{20} $& 88  & 2.1 & 67 & 256&27\\
M & 12$^\mathrm{h}$12$^\mathrm{m}$35.9$^\mathrm{s}$& $-$53$\degr$45$\arcmin$18.0$\arcsec$  & 253.3 &  4.2 & 379 & 21  & 10.8 & 0.06 & $1.4\cdot10^{20} $&  96 & 2.2 & 44 & 821\\
O & 12$^\mathrm{h}$12$^\mathrm{m}$30.5$^\mathrm{s}$& $-$53$\degr$43$\arcmin$26.0$\arcsec$  & 252.6 &  3.3 & 237 & 20  & 10.2 & 0.08 & $9.7\cdot10^{19 }$& 112 & 1.3 & 26 & 302\\
I & 12$^\mathrm{h}$12$^\mathrm{m}$36.8$^\mathrm{s}$& $-$53$\degr$42$\arcmin$14.0$\arcsec$  & 253.7 &  3.9 & 332 & 26  & 13.2 & 0.08 & $1.5\cdot10^{20} $&  80 & 2.8 & 75 & 928\\
K & 12$^\mathrm{h}$12$^\mathrm{m}$38.6$^\mathrm{s}$& $-$53$\degr$38$\arcmin$45.9$\arcsec$  & 252.9 &  3.4 & 249 & 29  & 14.5 & 0.11 & $1.7\cdot10^{20 }$& 112 & 2.2 & 62 & 544&19\\
N & 12$^\mathrm{h}$12$^\mathrm{m}$34.1$^\mathrm{s}$& $-$53$\degr$40$\arcmin$30.0$\arcsec$  & 253.4 &  3.9 & 325 & 24  & 12.1 & 0.07 & $1.5\cdot10^{20 }$& 104 & 2.1 & 56 & 677\\
C & 12$^\mathrm{h}$13$^\mathrm{m}$15.4$^\mathrm{s}$& $-$53$\degr$34$\arcmin$36.9$\arcsec$  & 253.7 &  2.4 & 127 & 22  & 11.3 & 0.18 & $9.7\cdot10^{19 }$& 136 & 1.0 & 20 & 133&29\\
D & 12$^\mathrm{h}$13$^\mathrm{m}$16.4$^\mathrm{s}$& $-$53$\degr$41$\arcmin$41.3$\arcsec$  & 252.7 &  2.7 & 159 & 20  & 10.3 & 0.13 & $9.1\cdot10^{19} $&  96 & 1.4 & 23 & 222\\
F & 12$^\mathrm{h}$13$^\mathrm{m}$20.1$^\mathrm{s}$& $-$53$\degr$43$\arcmin$25.2$\arcsec$  & 252.6 &  3.2 & 216 & 16  &  8.2 & 0.07 & $8.7\cdot10^{19} $&  80 & 1.6 & 19 & 345\\
G & 12$^\mathrm{h}$13$^\mathrm{m}$12.0$^\mathrm{s}$& $-$53$\degr$44$\arcmin$29.4$\arcsec$  & 252.2 &  3.1 & 204 & 19  &  9.4 & 0.09 & $1.0\cdot10^{20} $&  88 & 1.8 & 26 & 358\\
P & 12$^\mathrm{h}$13$^\mathrm{m}$10.2$^\mathrm{s}$& $-$53$\degr$46$\arcmin$21.4$\arcsec$  & 252.1 &  2.0 &  83 & 16  &  8.0 & 0.19 & $5.6\cdot10^{19} $&  88 & 0.9 & 11 &  77&37\\
Q & 12$^\mathrm{h}$13$^\mathrm{m}$08.3$^\mathrm{s}$& $-$53$\degr$39$\arcmin$25.5$\arcsec$ & 252.5 &  2.4 & 122 & 27   & 13.8 & 0.22 & $1.3\cdot10^{20} $& 144 & 1.3 & 31 & 158\\
H & 12$^\mathrm{h}$12$^\mathrm{m}$48.5$^\mathrm{s}$& $-$53$\degr$41$\arcmin$57.9$\arcsec$   & 253.5 &  2.5 & 131 & 19 &  9.8 & 0.15 & $8.5\cdot10^{19} $&  88 & 1.4 & 24 & 185\\
L & 12$^\mathrm{h}$12$^\mathrm{m}$55.7$^\mathrm{s}$& $-$53$\degr$41$\arcmin$29.8$\arcsec$  & 253.4 &  2.8 & 176 & 16  &  8.1 & 0.09 & $8.3\cdot10^{19} $&  80 & 1.5 & 19 & 267\\
R & 12$^\mathrm{h}$13$^\mathrm{m}$13.0$^\mathrm{s}$& $-$53$\degr$50$\arcmin$21.1$\arcsec$  & 252.4 &  2.4 & 130 & 14  &  7.0 & 0.11 & $6.2\cdot10^{19 }$&  64 & 1.4 & 13 & 185\\
S & 12$^\mathrm{h}$13$^\mathrm{m}$03.9$^\mathrm{s}$& $-$53$\degr$47$\arcmin$41.5$\arcsec$ & 252.8 &  3.4 & 245 & 14   &  7.1 & 0.06 & $8.2\cdot10^{19} $&  96 & 1.3 & 15 & 307\\
T & 12$^\mathrm{h}$12$^\mathrm{m}$49.4$^\mathrm{s}$& $-$53$\degr$37$\arcmin$33.4$\arcsec$  & 252.8 &  3.5 & 262 & 49  & 24.6 & 0.19 & $3.0\cdot10^{20} $& 152 & 2.9 & 138& 752&22\\[1ex] \hline

\end{tabular}
\end{table*}

\subsection{HVC\,291+26+195}\label{hvc2}

\subsubsection{Parkes data}

\begin{figure*}
\includegraphics[width=\textwidth,clip]{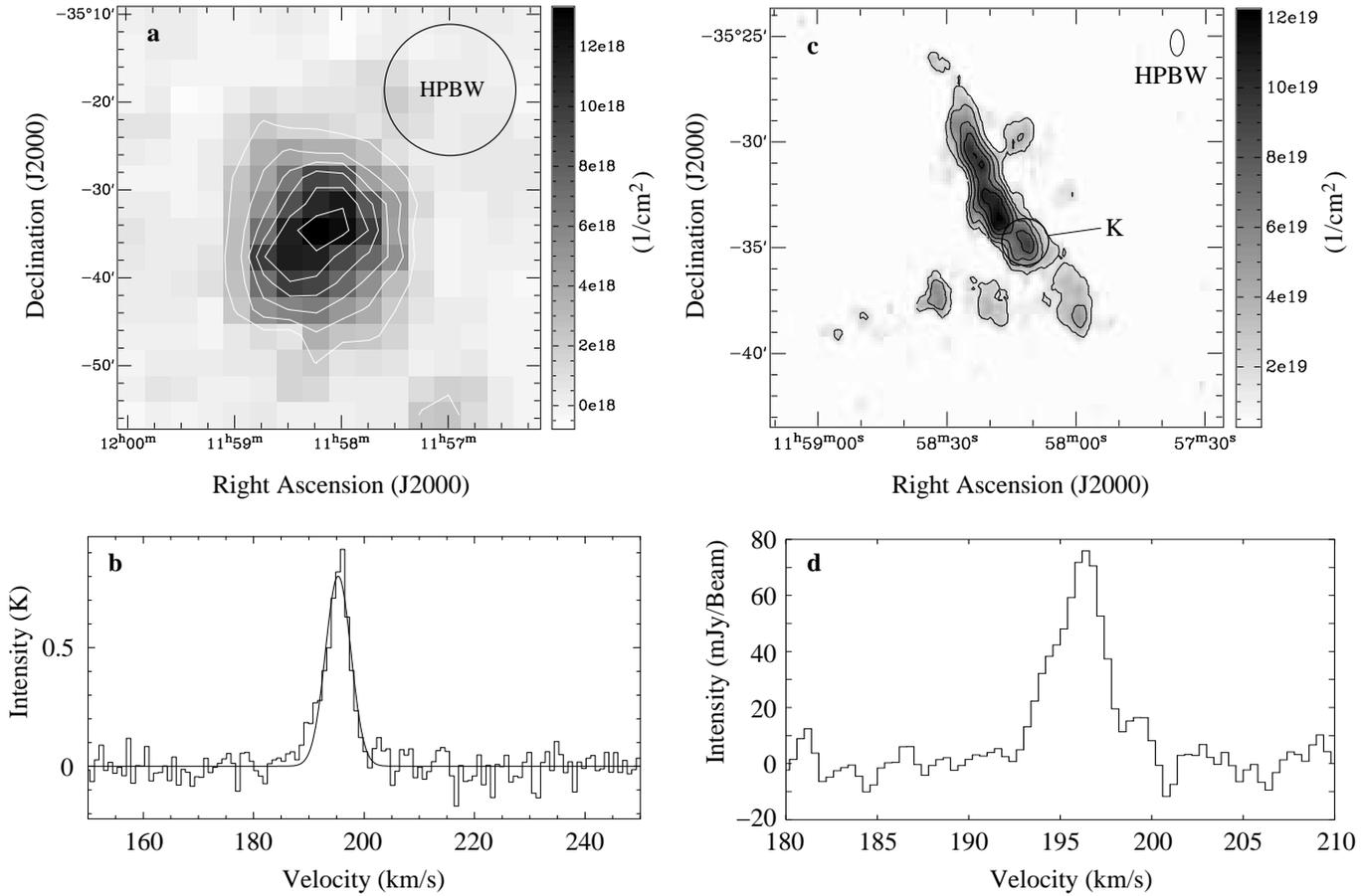}
\caption{\textbf{a}: \ion{H}{i} column density map of HVC\,291+26+195  observed with Parkes. The contours range from
$2 \cdot 10^{18}$ to $1.2 \cdot 10^{19}$ cm$^{-2}$ in steps of $2 \cdot 10^{18}$ cm$^{-2}$.
\textbf{b}: Average spectrum of all 12 positions in the direction of HVC~291+26+195 in which the spectral line was detected
with at least $3 \sigma$. The Gaussian fit (solid line) indicates that only a single line component is present.
\textbf{c}: Column density map of HVC~291+26+195  observed with the ATCA. The contours range from $2 \cdot 10^{19}$ to
$1.2 \cdot 10^{20}$ cm$^{-2}$ in steps of $2 \cdot 10^{19}$ cm$^{-2}$. The ATCA beam of $74.6\arcsec \times 38.4\arcsec$ HPBW
is also shown. \textbf{d}: A typical spectrum of clump K.}
\label{fig_hvc2_new}
\end{figure*}

Fig.~\ref{fig_hvc2_new}a shows the column density map of HVC\,291+26+195. Apparently, the Parkes beam of
14$\farcm$1 HPBW does not resolve the HVC. The investigation of the line profiles reveals that there is
only one line component in the spectra of this cloud. In Fig.~\ref{fig_hvc2_new}b we plotted the average
of all 12 spectra of HVC\,291+26+195. The line width of the component is about 5 \textrm{km\,s}$^{-1}$  FWHM.
The resulting upper temperature limit is $T_\mathrm{max}=640$ K. The data show no evidence for a warm gas phase.
The peak column density is $N_\mathrm{HI}=(1.3 \pm 0.1) \cdot 10^{19}$ cm$^{-2}$. The resulting parameters from
the Gaussian fits are similar to the results for the cold component of HVC\,297+09+253.

\subsubsection{ATCA data}

Figure \ref{fig_hvc2_new}c shows a column density map of HVC\,291+26+195  observed with the ATCA.
The column density distribution of the HVC is elongated containing several unresolved concentrations. The main
filament is accompanied by a few small clouds. The peak column densities of the individual clumps range from $3 \cdot 10^{19}$ cm$^{-2}$
to $1 \cdot 10^{20}$ cm$^{-2}$.

The total \ion{H}{i} mass of HVC\,291+26+195 is $M_\mathrm{HI}$ = $5.9 \cdot 10^3~\textit{M}_{\sun}\,[d^2/50\,\mathrm{kpc}]$
and $M_\ion{H}{i}$ = $5.6 \cdot 10^3~ \textit{M}_{\sun}\,[d^2/50\,\mathrm{kpc}]$,
measured with Parkes and ATCA respectively.
The ATCA detects $95 \%$ of the mass which was measured with Parkes, confirming that not much diffuse gas is
present in HVC\,291+26+195.

The \ion{H}{i} lines of HVC\,291+26+195 are quite narrow ($\Delta v_\mathrm{FWHM}=2\ldots4$ $\mathrm{km \, s}^{-1}$). As an example Fig.~\ref{fig_hvc2_new}d
shows the spectrum of clump K that has a line width of $\Delta v_{\rm FWHM} =$ 3.5\,\textrm{km\,s}$^{-1}$.
The line widths of all clumps are between $2 \ldots 4$ \textrm{km\,s}$^{-1}$. Assuming a distance of $d=50$ kpc (see Sect. \ref{distance} and \ref{LeadingArm}), the upper limit for the
pressure is  $(p / \rm{k_\mathrm{B}})_\mathrm{max} = 378 \pm 67$~K~cm$^{-3}~[d/\,50~ \mathrm{kpc}]^{-1}$ for clump K and
$(p / \rm{k_\mathrm{B}})_\mathrm{max} = 154 \pm 43$~K~cm$^{-3}~[d/\,50~ \mathrm{kpc}]^{-1}$ for clump N of HVC\,291+26+195. The important parameters of all investigated
clumps of HVC\,291+26+195 are listed in Table~\ref{tab_hvc2_atca}.

In summary, the ATCA observations of HVC\,291+26+195 reveal, as expected, only a cold gas phase having physical
properties similar to the cold phase of HVC\,297+09+253.

\begin{table*}
\caption{Important physical parameters for all clumps of HVC 291+26+195. For all distance-dependent parameters a distance of 50 kpc is assumed. The columns give the name of the clump, the coordinates, the velocity in the LSR frame, the line width $\Delta v_\mathrm{FWHM}$, the lower and upper temperature limits $T_\mathrm{min}$ and $T_\mathrm{max}$, the SNR, the optical depth $\tau$, the column density $N_\mathrm{HI}$, the angular diameter $\phi$, the particle density $n$, and the lower $\left(\frac{p}{k_\mathrm{B}}\right)_\mathrm{min}$ and upper pressure limit $\left(\frac{p}{k_\mathrm{B}}\right)_\mathrm{max}$ as well as the estimated distances $d$.}
\label{tab_hvc2_atca}
\tiny
\centering
\begin{tabular}{ccccccccccccccc}
\hline\hline
\rule{0pt}{3ex}clump&RA&Dec&$v_\mathrm{LSR}$&$\Delta v_\mathrm{FWHM}$&$T_\mathrm{max}$&$T_\mathrm{min}$&SNR&$\tau_\mathrm{min}$&$N_\mathrm{HI}$&$\phi$&$n$&$\left(\frac{p}{k}\right)_\mathrm{min}$&$\left(\frac{p}{k}\right)_\mathrm{max}$&$d$\\

\rule{0pt}{3ex}&&&[km\,s$^{-1}$]&[km\,s$^{-1}$]&[K]&[K]&&&[cm$^{-2}$]&[$\arcsec$]&[cm$^{-3}$]&[cm$^{-3}$K]&[cm$^{-3}$K]&[kpc]\\[1ex]\hline
\rule{0pt}{3ex}

A & 11$^\mathrm{h}$58$^\mathrm{m}$31.7$^\mathrm{s}$ & $-$35$\degr$37$\arcmin$15.7$\arcsec$ & 192.1  & 3.4 & 6.9 & 251 &  10  & 0.04 & $6.0\cdot10^{19}$  &  72 & 1.2  &  12  &  307&16\\
B & 11$^\mathrm{h}$58$^\mathrm{m}$48.8$^\mathrm{s}$ & $-$35$\degr$38$\arcmin$26.9$\arcsec$ & 192.0  & 2.5 & 5.7 & 137 &   8  & 0.06 & $3.8\cdot10^{19}$  &  80 & 0.7  &  6  &  96\\
C & 11$^\mathrm{h}$58$^\mathrm{m}$35.7$^\mathrm{s}$ & $-$35$\degr$37$\arcmin$39.5$\arcsec$ & 191.2  & 3.0 & 5.3 & 195 &   8  & 0.04 & $3.8\cdot10^{19}$  &  88 & 0.6  &  5  & 124\\
D & 11$^\mathrm{h}$58$^\mathrm{m}$17.3$^\mathrm{s}$ & $-$35$\degr$38$\arcmin$19.8$\arcsec$ & 194.9  & 2.3 & 6.0 & 112 &   9  & 0.08 & $3.6\cdot10^{19}$  &  96 & 0.6  &  5  &  63&35\\
E & 11$^\mathrm{h}$58$^\mathrm{m}$21.2$^\mathrm{s}$ & $-$35$\degr$37$\arcmin$47.9$\arcsec$ & 194.5  & 3.5 & 5.2 & 272 &   8  & 0.03 & $4.0\cdot10^{19}$  & 120 & 0.5  &  4  & 134\\
F & 11$^\mathrm{h}$58$^\mathrm{m}$18.6$^\mathrm{s}$ & $-$35$\degr$37$\arcmin$07.9$\arcsec$ & 194.0  & 2.7 & 5.2 & 155 &   8  & 0.05 & $2.9\cdot10^{19}$  & 112 & 0.4  &  3  &  59\\
G & 11$^\mathrm{h}$58$^\mathrm{m}$54.7$^\mathrm{s}$ & $-$35$\degr$38$\arcmin$58.5$\arcsec$ & 193.7  & 3.8 & 4.7 & 310 &   7  & 0.02 & $5.3\cdot10^{19}$  &  54 & 1.4  & 10  & 445\\
H & 11$^\mathrm{h}$58$^\mathrm{m}$33.0$^\mathrm{s}$ & $-$35$\degr$26$\arcmin$02.8$\arcsec$ & 195.9  & 2.5 & 5.3 & 136 &   8  & 0.06 & $3.9\cdot10^{19}$  &  88 & 0.7  &  5  &  90\\
I & 11$^\mathrm{h}$58$^\mathrm{m}$12.0$^\mathrm{s}$ & $-$35$\degr$29$\arcmin$39.7$\arcsec$ & 196.0  & 2.8 & 6.7 & 173 &  10  & 0.06 & $4.4\cdot10^{19}$  & 104 & 0.6  &  6  & 107\\
J & 11$^\mathrm{h}$58$^\mathrm{m}$03.5$^\mathrm{s}$ & $-$35$\degr$32$\arcmin$35.9$\arcsec$ & 195.7  & 1.4 & 4.9 &  45 &   7  & 0.16 & $1.7\cdot10^{19}$  &  96 & 0.3  &  2  &  12\\
K & 11$^\mathrm{h}$58$^\mathrm{m}$10.1$^\mathrm{s}$ & $-$35$\degr$34$\arcmin$60.0$\arcsec$ & 196.1  & 3.5 &10.8 & 272 &  16  & 0.06 & $9.8\cdot10^{19}$  & 104 & 1.4  & 22  & 378& 13\\
L & 11$^\mathrm{h}$58$^\mathrm{m}$02.8$^\mathrm{s}$ & $-$35$\degr$36$\arcmin$35.9$\arcsec$ & 196.2  & 2.5 & 7.3 & 136 &  11  & 0.08 & $4.3\cdot10^{19}$  &  96 & 0.7  &  7  &  91\\
M & 11$^\mathrm{h}$57$^\mathrm{m}$59.6$^\mathrm{s}$ & $-$35$\degr$38$\arcmin$11.7$\arcsec$ & 197.3  & 2.3 & 8.4 & 112 &  12  & 0.11 & $5.5\cdot10^{19}$  & 120 & 0.7  &  8  &  76\\
N & 11$^\mathrm{h}$58$^\mathrm{m}$27.8$^\mathrm{s}$ & $-$35$\degr$29$\arcmin$15.5$\arcsec$ & 195.3  & 2.2 &10.0 & 106 &  15  & 0.14 & $6.3\cdot10^{19}$  &  64 & 1.4  & 21  & 154&25\\
O & 11$^\mathrm{h}$58$^\mathrm{m}$23.8$^\mathrm{s}$ & $-$35$\degr$30$\arcmin$35.7$\arcsec$ & 195.4  & 3.0 &13.6 & 198 &  20  & 0.10 & $1.1\cdot10^{20}$  & 104 & 1.6  & 32  & 321\\
P & 11$^\mathrm{h}$58$^\mathrm{m}$21.2$^\mathrm{s}$ & $-$35$\degr$32$\arcmin$19.9$\arcsec$ & 195.3  & 3.1 &13.4 & 212 &  20  & 0.09 & $1.1\cdot10^{20}$  &  64 & 2.5  & 49  & 532\\
Q & 11$^\mathrm{h}$58$^\mathrm{m}$18.6$^\mathrm{s}$ & $-$35$\degr$33$\arcmin$48.0$\arcsec$ & 195.2  & 3.2 &13.2 & 227 &  19  & 0.08 & $1.2\cdot10^{20}$  &  64 & 2.8  & 54  & 634\\[1ex] \hline

\end{tabular}
\end{table*}

\section{Discussion}\label{disscussion}

\subsection{Evidence for ram-pressure interaction}

The \ion{H}{i} data of HVC\,297+09+253 show a distinct head-tail structure. The high column density region, the
head of the cloud, contains both a narrow and a broad component. In the northern-most part of the CHVC only a
narrow line component is detected, while there is solely a broad component in the tail (see Sect.~\ref{hvc1}).

The observed morphology suggests a ram-pressure interaction with an ambient medium, where the warm gas phase is stripped off the CHVC due to friction forces \citep{bruenskerpsmith00,bruenskerppagels01}.
The clumps may have formed due to compression of the gas at the leading edge of an CHVC and local density fluctuations in combination with cooling processes.

The existence of a two-component gas structure is widely observed among HVCs and CHVCs \citep{Cram76, wolfiremckeehollenbachtielens95, burtonbraunchengalur01, deheijbraunburton02, braunburton99, westmeier05}. The 2D hydrostatic simulations of CHVCs by \citet{sternbergmckeewolfire02} are consistent with these observations. Neutral atomic hydrogen clouds can accordingly show a stable multiphase structure in an environment with conditions comparable to those in the outer Galactic halo. The hydrodynamical simulations of \citet{quilismoore01} and \citet{vieserhensler01} of HVCs moving through a diffuse medium show
that the lowest temperatures of interacting HVCs are found in the head of the clouds whereas higher
temperatures occur in the tail of HVCs where the material was stripped off the main body of the cloud. \citet{Konz} show in their hydrodynamical simulations that the cold gas in the head of the cloud remains unaffected by the stripping process. This is qualitatively consistent with our observations of HVC\,297+09+253.
The smallest line widths of $\Delta v_\mathrm{FWHM} \approx 2$ \textrm{km\,s}$^{-1}$ (indicating the lowest
temperatures) are located in the head of HVC\,297+09+253 while the broadest lines of about
$\Delta v_\mathrm{FWHM}=20$ \textrm{km\,s}$^{-1}$ are observed in the tail.

The tail of HVC\,297+09+253 shows higher radial velocities than the high column density head (Fig. \ref{hvc297composite}f).
\cite{bruenskerpsmith00} analysed 252 HVCs. $20 \%$ of the clouds show a head-tail structure combined with a gradient in
column density and in velocity along the major axis of the clouds. But none of the HVCs in this sample shows an increasing
velocity in the direction of the tail.

Part of this gradient can be explained as a projection effect from the solar velocity vector. The gradient decreases slightly
from $\Delta v_\mathrm{LSR}=3$ \textrm{km\,s}$^{-1}$ to about $\Delta v_\mathrm{GSR}=2.5$ \textrm{km\,s}$^{-1}$ when
transforming into the Galactic standard-of-rest frame. The velocity vector of the HVC produces a similar effect. If we assume that the velocity is produced by a pure projection effect we would get extremely high spatial velocities of the order of $>500$ \textrm{km\,s}$^{-1}$, which is significantly larger than the maximum radial velocities observed for HVCs around the Milky Way.

The apparent contradiction is resolved by considering a slightly varying orientation of the velocity vector over the extent
the tail. If the total velocity of the HVC is for instance 200~\textrm{km\,s}$^{-1}$, a twist of only one degree
would be sufficient to explain the observed gradient. A slight deflection of the stripped matter in the tail could have
several reasons like an intrinsic velocity of the ambient medium, e.g. a Galactic wind or turbulence in the hot halo of the
Milky Way. Moreover, the Galactic magnetic field could influence the gas in the tail. \cite{Konz} performed
magnetohydrodynamical simulations of interacting HVCs and concluded that magnetic fields play an important role in
stabilising the head while forming a tail.

The cloud HVC\,291+26+195 shows cold compact clumps with very small line widths comparable to the head of
HVC\,297+09+253, but no diffuse component. This might be interpreted as a cloud in a later stage of interaction with the
ambient medium, where the diffuse warm component was already stripped off and evaporated within the hot gas phase of the Galactic halo \citep{quilismoore01} while the cold clumps are still neutral and observable in \ion{H}{i} emission.

\begin{figure}
\includegraphics[width=0.46\textwidth,clip]{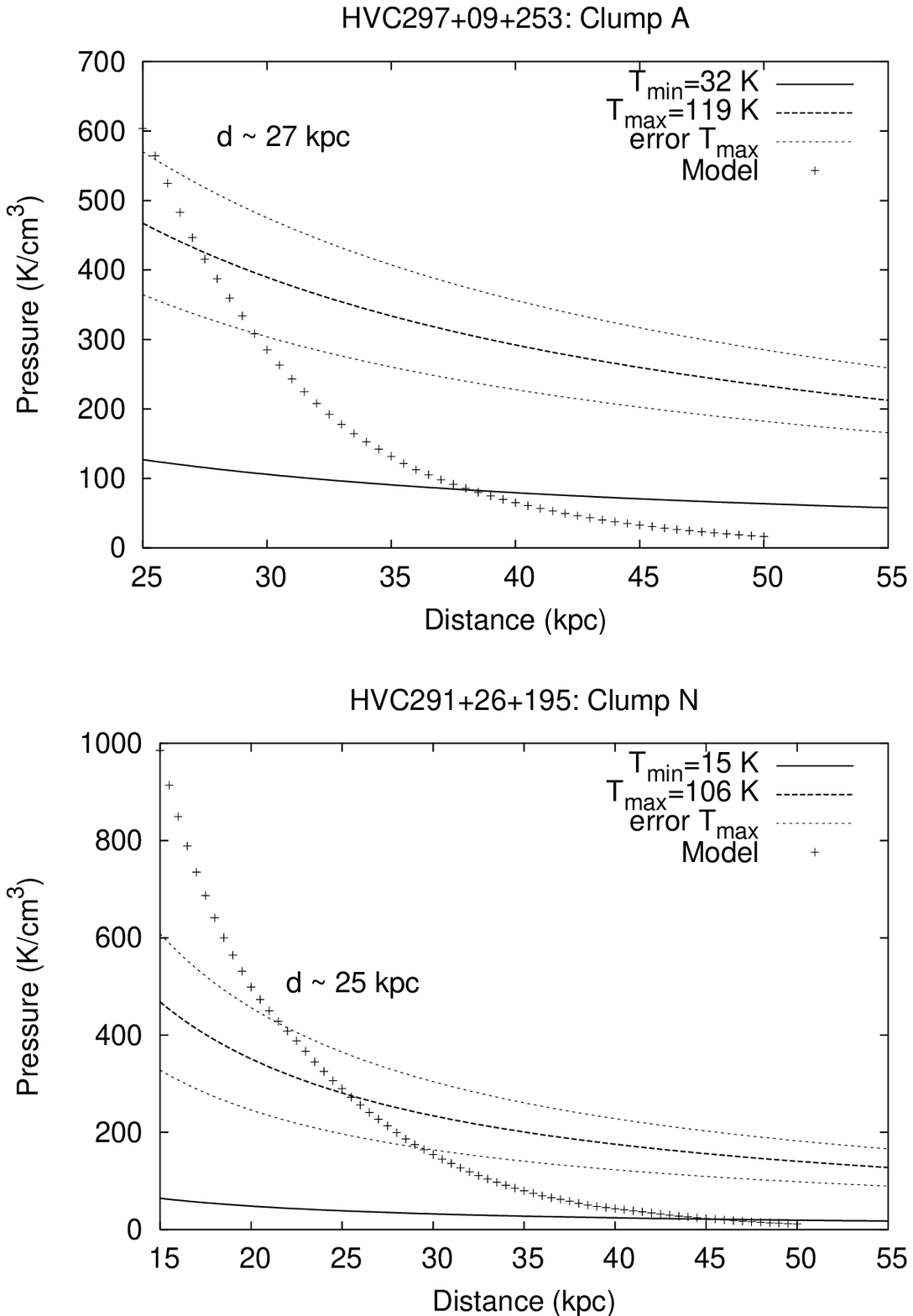}
\caption{Pressure variation as a function of distance and statistical error curves (dotted lines) for two clumps
of HVC~297+09+253 and HVC~291+26+195. Furthermore, the pressure variation of the Milky Way gas (crosses) derived
from a model developed by \citet{kalberla03} is shown. \textbf{Upper panel:} Clump~A of HVC~297+09+253. \textbf{Lower panel:} Clump~N
of HVC~291+26+195.}
\label{fig_pressure_clumps}
\end{figure}

\subsection{Distance estimate for the two CHVCs}\label{distance}

In Sect.~\ref{hvc1atca}, we estimated the pressure of the cold clumps of HVC\,297+09+253 and HVC\,291+26+195, assuming an ideal gas, a
spherically-symmetric cloud, and a constant particle density. The very small line widths allow to constrain
the kinetic temperature of the clumps and, thus, to estimate upper and lower pressure limits for all clumps in the two
CHVCs (see Tables \ref{tab_hvc_atca} and \ref{tab_hvc2_atca}) under the above assumptions.

We compare the pressure of some of the clumps in the two CHVCs as a function of distance with the pressure of the
surrounding medium derived from the Milky Way model of \citet{kalberla03}. Fig. \ref{fig_pressure_clumps}
shows the results for two exemplary clumps of HVC\,297+09+253  and HVC\,291+26+195.

The pressure of the clumps should be at least as high as the pressure of the surrounding medium. The intersection
point of the pressure curves for the Milky Way gas and the clumps therefore provides a distance estimate.
With the mentioned assumptions the corresponding distance estimates for HVC\,297+09+253 and HVC\,291+26+195 are in the range of $d\approx10 \ldots 60$ kpc (Table~1 and 2).

One should keep in mind, though, that the distribution of the pressure of the circum-/intergalactic medium is not well constrained by observations. There is observational evidence from absorption line measurements that the thermal pressure could be significant even at larger distances of several hundred kpc from the Galaxy \citep{sembachwakkeretal03}. Results from X-ray absorption data by \citet{rasmussenkahnpaerels03} indicate a substantial value of $P/\mathrm{k_\mathrm{B}} \approx 100~ \mathrm{cm}^{-3} \mathrm{K}$ beyond 100 kpc from the Galaxy. A higher value of the pressure would shift the pressure curve of \citet{kalberla03} slightly upwards. On the other hand, our calculated values for the internal pressure of some of the clumps are only lower limits because of their significant optical depth for which we cannot correct. Deviations from the previous assumptions (Sec. 3.1.2) add additional uncertainties to the calaculated parameters. Consequently, the determined distances should be considered rough estimates only.

\subsection{Evidence for an association of both CHVCs with the Leading Arm} \label{LeadingArm}

Fig.~\ref{fig_MS_NHImap} shows a column density map of the Magellanic System \citep{Bruenskerp05}, including the
Small and the Large Magellanic Cloud (SMC and LMC), the Magellanic Bridge, which connects the two Magellanic
Clouds, the Magellanic Stream (MS), and the Leading Arm (LA). These features were likely generated in an
interaction between the Magellanic Clouds with the Milky Way \citep{Yosnoguchi03}. The LA consists of many
clumpy filaments \citep{putmanetal.98} and shows a two-component gas structure \citep{Bruenskerp05}, comparable to the results derived
for HVC\,297+09+253 and HVC\,291+26+195.

The LA can be divided into three parts, the LA I, II, and III. In the south-eastern part of LA II the observed
velocities are between $265 \ldots 300$ \textrm{km\,s}$^{-1}$ in the LSR frame. Thus, HVC\,297+09+253  deviates
by only $10 \ldots 15$~km~s$^{-1}$  from the velocities of this region, which is in good agreement within the velocity dispersion of the gas in LA II of 10 \textrm{km\,s}$^{-1}$. HVC\,291+26+195  lies in the
velocity range of the northern part of the LA II ($190 \ldots 275$ \textrm{km\,s}$^{-1}$).
Numerical simulations of the Magellanic System, e.g. \citet{Yosnoguchi03}, predict distances for the gas in the
Leading Arm in the range $30 \ldots 60$~kpc. This distance range is consistent with the results of the distance
estimation for the two clouds described in Sect.~\ref{distance}.

The velocity vector of the LMC observed by \citet{Marel} indicates that the orbit of the Large
Magellanic Cloud is in the direction of increasing Galactic latitude. This is consistent with the observed
orientation of the tail of HVC\,297+09+253, which is perpendicular to the Galactic plane and pointing towards the LMC.

The proximity of the two clouds to the Leading Arm, the comparable velocities, the orientation of the tail and
the distance estimates make an association with the Magellanic System likely, implying a distance range of 10 kpc $\leq d \leq$ 60 kpc for the two HVCs.

\section{Summary and outlook}\label{summary}

We analysed single-dish (Parkes) and interferometer (ATCA) data of two compact high-velocity clouds located in the vicinity of the Leading Arm of the Magellanic System. The
observations with both telescopes allow us to investigate the total \ion{H}{i} mass and the extended structure
as well as the small-scale structure within the clouds.

The analysis of HVC\,297+09+253 reveals that the cloud has a two-component gas structure. The cold and the warm
gas phase are partly spatially separated. Cold compact clumps are embedded in a diffuse warm \ion{H}{i} gas.
Both the gradient in line width and in column density as well as the morphological asymmetry show that this
cloud reveals a head-tail structure. The presence of the cold clumps and the head-tail structure can be
explained by an interaction of this cloud with an ambient medium. All clumps show very narrow lines of
$\Delta v_\mathrm{FWHM} < 4$~\textrm{km\,s}$^{-1}$, allowing us to constrain physical parameters like the
temperature and the pressure of the gas.

In the case of HVC\,297+09+253 we also find cold compact clumps with very narrow lines but no diffuse, warm gas.
Possibly, these clumps were also generated in an interaction with an ambient medium. The diffuse gas possibly
was already stripped off the cold core due to friction forces.

We presented a method to estimate distances for the two CHVCs by comparing the pressure of the
clumps in the CHVCs as a function of distance with the pressure of the surrounding Milky Way gas. The distances estimated for all clumps are of the order of $10 \ldots 60$~kpc.

All our results provide evidence for an association of the two CHVCs with the Leading Arm of the Magellanic System
at a distance of $d \approx 10 \ldots 60$ kpc. This distance estimation allows us to constrain the physical parameters of the clouds and to learn more about the physical conditions in the environment of the Leading Arm.

To learn more about the nature of HVCs in general, it is necessary to make absorption measurements towards suitable background sources at known distance. This would provide a direct distance
estimate as well as information about the metallicity of the clouds to confirm an association with
the Magellanic Clouds. High-resolution observations of CHVCs in the vicinity of the Leading Arm and the Magellanic Stream are necessary to resolve other small-scale structures for a statistical analysis of core-halo structures in CHVCs. Furthermore, H$\alpha$ measurements will be of great importance to investigate the
interaction with an ambient medium in considerably more detail. Simulations are necessary for a better
understanding of how such cold clumps are generated and to analyse the conditions under which interactions occur.

\begin{acknowledgements}
Many thanks to Philipp Richter for his helpful comments. Thanks to Benjamin Winkel for his helping hands.
T.W. is supported by the DFG through grant KE 757/4-1.
\end{acknowledgements}
\bibliographystyle{aa}
\bibliography{references}

\end{document}